\documentclass[twocolumn,superscriptaddress,floatfix,showpacs,amssymb,amsmath,amsfonts,aps]{revtex4}
\date{\today }
\input{epsf}
\usepackage{epsfig}
\usepackage{latexsym}
\usepackage{amssymb}
\usepackage{graphicx}
\usepackage[dvips]{color}
\begin{document}  
\title{\bf Evidence for the Strong Dominance of Proton-Neutron
Correlations in Nuclei}
\author{E. Piasetzky}
\affiliation{School of Physics and Astronomy, Sackler Faculty of Exact
Science,  
Tel Aviv  University, Tel Aviv 69978, Israel}
\author{M. Sargsian}
\affiliation{Department of Physics, Florida International University,
Miami, FL 33199, U.S.A}
\author{L. Frankfurt}
\affiliation{School of Physics and Astronomy, Sackler Faculty of Exact
Science,  
Tel Aviv  University, Tel Aviv 69978, Israel}
\author{M. Strikman}
\affiliation{Department of Physics, The Pennsylvania State University,
University Park, PA, U.S.A}
\author{J. W.  Watson}
\affiliation{Department of Physics, Kent State University,Kent, OH
44242,U.S.A}
\date{\today}

\begin{abstract}
{\bf Abstract:} We analyze recent data from  high-momentum-transfer 
$(p,pp)$ and $(p,ppn)$  reactions on Carbon. For this analysis, 
the two-nucleon short-range correlation~(NN-SRC) model for backward 
nucleon emission is extended to include the motion of the NN-pair in the 
mean field. The model is found to describe major characteristics of the
data. 
Our analysis demonstrates  that the removal of a proton from the nucleus 
with initial momentum  $275-550$~MeV/c  is $92^{+8}_{-18}\%$ of the time 
accompanied by the emission of a correlated neutron that carries momentum 
roughly  equal and opposite to the initial proton momentum. 
Within the NN-SRC dominance assumption the data indicate that the 
probabilities of $pp$ or $nn$  SRCs in the nucleus are at least a factor 
of six smaller than that of $pn$ SRCs. Our result is the first estimate 
of the isospin structure of NN-SRCs in nuclei, and may have important 
implication for modeling the  equation of state of asymmetric nuclear
matter.
\end{abstract}
\maketitle
\bibliographystyle{unsrt}

Studies of short-range nucleon correlations~(SRCs) in nuclei are important 
for understanding the short-distance and large-momentum properties of
nuclear
ground state wave functions. The relevant 
distances 
in two-nucleon~(NN)-SRCs are expected to be comparable to that in 
neutron stars  corresponding to 4-10 times the central density of 
nuclei \cite{hnm}. 
Thus SRC studies are essential in understanding the structure of 
cold dense nuclear matter.
In this context the isospin content 
of SRCs (i.e. $pn$ vs. $pp$ and $nn$ pairs) is important for
understanding the 
structure of nuclear matter made of either protons or neutrons.  
Studies of SRCs also give the best hope of understanding the nature of the 
short-range NN repulsion.

SRCs in nuclei have been actively investigated for over three decades 
(see e.g.\cite{PSH97}). 
However, experimental studies of the microscopic structure of SRCs 
were largely restricted  due to moderate momentum-transfer 
kinematics in which it is difficult to resolve SRCs.
Recently, several experiments \cite{Weinstein,Kim1,Kim2,Eip1,Eip2} made 
noticeable progress in  understanding dynamical aspects of SRCs.
For $Q^2>1$~GeV$^2$, Refs~\cite{Kim1,Kim2} observed 
Bjorken $x_B$ scaling  for ratios of inclusive $(e,e')$ cross sections  
of nuclei $A$ to the $^3He$ nucleus when $x_B\ge 1.4$.
This confirms the earlier observation of scaling for nucleus-to-deuteron 
cross section ratios\cite{FSreps,Day}, and 
indicates directly that the electrons probe high-momentum bound nucleons 
coming from local sources in nuclei (i.e. SRCs) with properties generally 
independent of the non-correlated residual nucleus.

Based on the NN-SRC picture, which is expected to dominate the
internal momentum range of $\sim 250-600$~MeV/c, one predicts a strong 
directional (back-to-back) correlation between the struck nucleon and its 
spectator in the SRC.  Experiments\cite{Weinstein,Eip1,Eip2} 
measured triple-coincidence events for the $^3He(e,e'pp)X$ and 
$^{12}C(p,ppn)X$ 
reactions, and clearly demonstrated the existence of such directional 
correlations. They also revealed a noticeable momentum 
distribution of the center of mass~(c.m.) of the NN-SRCs.

In this work we extend the NN-SRC model used in the analyses of $A(p,pp)X$
data\cite{Yaron}, to incorporate the effects of the c.m.  motion of SRCs. 
This allows us to estimate  the  probability for correlated  
neutron emission following removal of a fast proton from the nucleus in 
$(p,ppn)$ reactions. 
Based on this model we extract from the data an upper limit to the 
relative probabilities of $pp$ and $nn$ vs $pn$ SRCs in $^{12}C$.

The measurements of $^{12}C(p,ppn)X$ reactions\cite{Eip1,Eip2} were 
performed with  the EVA spectrometer 
at the AGS accelerator at Brookhaven National Laboratory \cite{EVA1,EVA2}.
EVA consists of a $0.8$~T superconducting solenoid, $3.3$~m long and 
$2$~m in diameter. The $5.9-9.0$~GeV/c proton beam was incident along the 
central axis. Coincident pairs of high transverse-momentum protons were 
detected with four concentric cylinders of straw tube chambers. The 
experimental kinematics are  discussed in more details 
later.   Neutrons were detected in coincidence with the 
quasi-elastic knockout of protons from $^{12}C$. The large momentum 
transfers $-t\ge 6 \, GeV^2$ in these processes greatly 
improve the resolving power of the probe and validate the instantaneous 
approximation for description of the removal of fast bound proton in 
the $pp\rightarrow pp$ subprocess. For each $(p,pp)$ event, 
the momentum of the struck proton~$\vec p_2$ 
before the reaction was reconstructed and compared (event by event) with
the 
measured coincident neutron momentum~$\vec p_n$.
Due to the $\sim s^{-10}$ dependence of the underlying hard 
$pp\rightarrow pp$ cross section, the scattering takes place preferentially 
off a bound proton  with large~$|p_2|$ in the direction of the beam 
(minimizing $s$)\cite{Farrar}, and hence should lead to a significant 
rate of emission of backward correlated nucleons due to scattering off 
NN-SRCs. Data confirming these characteristics of  $A(p,ppn)X$ reactions
are 
shown in  Fig.~1 for $^{12}C$. The data show  no directional correlation
for 
neutrons with $|p_n|$ below the Fermi sea level ($k_F=220$~MeV/c). 
Above $k_F$ a strong back-to-back directional correlation between 
$\vec p_2$ and $\vec p_n$ is evident. 
\begin{figure}[ht]
\vspace{0.6cm}
\centerline{\epsfig{height=3.6cm,width=8cm,file=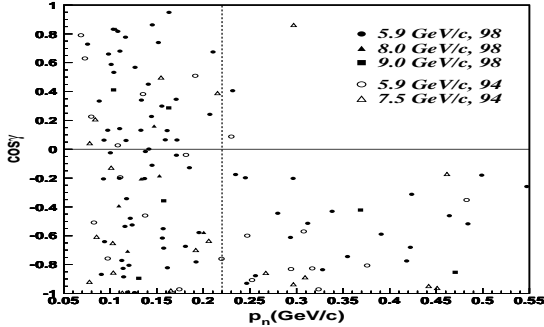}}
\vspace{-0.4cm}
\caption{ The correlation between~$p_n$ and its direction $\gamma$ relative 
to $\vec p_2$. Data labeled by $94$ 
and $98$ are from Ref.\cite{Eip1} and \cite{Eip2} respectively. The momenta
are the beam momenta.  The dotted vertical line corresponds to 
$k_F=220$~MeV/c.}
\end{figure}

In Ref.\cite{Eip2} the large value of the following ratio:     
\begin{equation}
F = \frac{\mbox{Number of (p,ppn) events ($p_2,p_n>k_F$)}}
{\mbox{Number of (p,pp) events ($p_2> k_F$)}}, 
\label{F}
\end{equation} 
was extracted, which indicates  that
in  the $250-550$~MeV/c region NN-SRCs must be the major source of
nucleons in 
nuclei. Within the SRC model, the numerator of $F$  is due to scattering 
off the $pn$-SRC while the denominator is due to scattering off 
any possible configuration ``pX'' which contains a proton with 
$|p_2|\sim 250-550$~MeV/c. All configurations such as 
$pn$- and $pp$- SRCs, the high-momentum tail of the mean-field  proton 
distribution, 
three and more nucleon SRCs, as well as SRCs containing non-nucleonic
degrees 
of freedom 
could contribute to ``pX''.  We define  $P_{pn/pX}$ as the
relative probability of finding a $pn$-SRC in the ``pX'' configuration.
In this work, using a theoretical description of the
$A(p,pp)X$ and $A(p,ppn)X$ reactions, we evaluate $P_{pn/pX}$
from  the above measured ratio $F$.

The present theoretical description of the $A(p,ppn)X$ reaction, in which 
the hard $pp\rightarrow pp$ subprocess is accompanied by the emission of 
a recoil ($k_F <p_n < 550$~MeV/c) neutron, 
is based on the light-cone distorted-wave  impulse approximation~(LC-DWIA).
This is an appropriate approximation for high-momentum-transfer reactions  
aimed at studies of the ground state properties of  nuclei since in 
this case the LC  momentum fraction $\alpha$~(defined below) 
of the nuclear constituents  
is approximately conserved without much distortion due to soft initial 
and final state interactions in  the reaction \cite{MMS01,FSS97}.
This validates the factorization of the hard $pp\rightarrow pp$ 
scattering from the soft reinteractions, allowing to express the 
$A(p,ppn)X$ cross section through the product of the $pp\rightarrow pp$
scattering  cross section off 
the bound proton, ${d\sigma/dt}$,  and the nuclear decay
function,~$D^{pn}$, 
as follows:
\begin{eqnarray}
\sigma^{p,ppn}= \sum\limits_{Z} K
{d\sigma^{pp}\over dt}(s,t)
{2D^{pn}(\alpha,\vec p_t,\alpha_n,\vec p_{tn}, P_{R+})\over \alpha}T_{ppn},
\label{p2pn}
\end{eqnarray}
where $\sigma^{p,ppn} \equiv  {d^{9}\sigma/ {d^3p_3\over E_3} 
{d^3p_4\over E_4}{d\alpha_n\over \alpha_n}d^2p_{tn}}$ and $K = {2\over \pi}
\sqrt{s^2-4m^2s}$.
Here $p_1$, $p_3$ and $p_4$ are incoming, scattered and knocked-out proton 
four-momenta and $p_2 = p_3+p_4 - p_1$. Also, $s=(p_3+p_4)^2$ and 
$t = (p_1-p_3)^2$. The function $D^{pn}$ 
represents  the joint probability of finding a proton in the nucleus with 
$\alpha = A{E_2-p^z_{2}\over E_{A}-P^z_A}$ and transverse momentum 
$\vec p_t$, and a 
recoil neutron in the residual $A-1$ nucleus with $\alpha_n$ 
and $\vec p_{tn}$. $P_{R+}=E_R-p^z_2$ is the LC ``plus'' component 
of the $A-1$ residual nuclear state. 
Here, $z\parallel \vec p_1$.
$T_{ppn}$ is the nuclear transparency for the 
yield of two fast protons and recoil neutron.  Eq.(\ref{p2pn}) 
correctly reproduced the angular correlation of Fig.~1 as well as average 
values of  
$\langle \alpha_n\rangle$ and $\langle cos(\gamma)\rangle$\cite{Eip1,Eip2}.

Within the  LC-DWIA, the $A(p,pp)X$ reaction is expressed through 
the spectral function, $S^p$,  as follows\cite{Yaron}:
\begin{eqnarray}
\sigma^{p,pp}= \sum\limits_{Z} K {d\sigma^{pp}\over dt}(s,t)
{2S^p(\alpha,\vec p_t,P_{R+})\over \alpha}T_{pp}
\label{p2p}
\end{eqnarray}
where $\sigma^{p,pp}\equiv  {d^{6}\sigma/ {d^3p_3\over E_3} {d^3p_4\over
E_4}}$, 
and $T_{pp}$ represents the nuclear transparency for the yield of two fast 
protons from the nucleus.  For SRCs we have:
\begin{eqnarray}
S^p(\alpha,\vec p_t,P_{R+}) = \sum\limits_{s}\int 
D^{ps}(\alpha,\vec p_t,\alpha_s,\vec p_{ts}, P_{R+}) 
{d\alpha_s\over \alpha_s}d^2p_{ts}
\label{d2s}
\end{eqnarray}
where the  summation is  over the possible types of the recoil
particles~$s$ from 
the SRC. 
The spectral function $S^p$ together with the mean-field contribution to
the 
spectral function is normalized to unity. 
At  small internal momenta, the LC-DWIA reduces smoothly to its 
nonrelativistic counterpart.  
Reasonable agreement is observed\cite{Yaron} 
in an extensive comparison of calculations based on Eq.(\ref{p2p}) with 
measured $\alpha$ and $p_t$ spectra\cite{Eip1}.  

The ratio $F$ defined in Eq.(\ref{F}) now  can be written as: 
\vspace{-0.2cm}
\begin{eqnarray}
F = {\int\limits_{\alpha^{min}}^{\alpha^{max}}
\int\limits_{p^{min}_t}^{p^{max}_t}
\int\limits_{\alpha^{min}_n}^{\alpha^{max}_n}
\int\limits_{p^{min}_{tn}}^{p^{max}_{tn}}
\sigma^{p,ppn} {d\alpha\over \alpha}d^2p_t {d\alpha_n\over \alpha_n}
d^2p_{tn}dP_{R+}\over 
\int\limits_{\alpha^{min}}^{\alpha^{max}}
\int\limits_{p^{min}_t}^{p^{max}_t}
\sigma^{p,pp} {d\alpha\over \alpha}d^2p_t dP_{R+}}
\label{Fth}
\end{eqnarray}
where the limits for 
integration are defined by the 
experimental conditions. 
The kinematics of the experiment\cite{Eip1,Eip2} lead to the integration 
by $P_{R+}$ that covers all the range relevant to quasielastic scattering.

Using Eq.(\ref{Fth}) and relations (\ref{p2pn}),(\ref{p2p}) and 
(\ref{d2s}),   
one can relate 
the above  defined $P_{pn/pX}$  averaged over the measured range of 
$\vec p_2$, $\vec p_n$ and $P_{R+}$ to the ratio-$F$ as follows.: 
\begin{equation}
P_{pn/pX} = \frac{F}{T_n R}\cdot
\label{PpnSRC}
\end{equation}
Here $T_n$ accounts for the attenuation of the neutron and 
\vspace{-0.2cm}
\begin{equation}
R \equiv {\int\limits_{\alpha^{min}}^{\alpha^{max}}
\int\limits_{p^{min}_t}^{p^{max}_t}
\int\limits_{\alpha^{min}_n}^{\alpha^{max}_n}
\int\limits_{p^{min}_{tn}}^{p^{max}_{tn}}
D^{pn} {d\alpha\over \alpha}d^2p_t {d\alpha_n\over \alpha_n}
d^2p_{tn}dP_{R+}\over 
\int\limits_{\alpha^{min}}^{\alpha^{max}}
\int\limits_{p^{min}_t}^{p^{max}_t}
S^{pn} {d\alpha\over \alpha}d^2p_t dP_{R+}},
\label{R}
\end{equation}
where $S^{pn}$ is the part of the spectral function $S^{p}$ related 
to $pn$-SRCs only.

In the derivation of Eq.(\ref{PpnSRC}) we used $T_{ppn}\approx T_{pp}T_{n}$,
which is justified for kinematics in which 
two energetic, $>3$~GeV/c protons are produced in the projectile 
fragmentation region while the neutron is detected with 
$|p_n| \le 550$~MeV/c in the backward direction. 
The accuracy of this approximation is proportional to the relative yield of 
recoil neutrons due to rescattering of the energetic protons off
uncorrelated 
neutrons in the nucleus, which amounts $\sim 5\%$ for the considered 
kinematics.

Note that Eq.(\ref{PpnSRC}) gives a well-defined upper limit for 
$P_{pn/pX}$, namely, if the nucleus is transparent to the recoil 
neutron ($T_n=1$), all the strength of the SRC is due to $pn$-SRCs,
and if the kinematic cuts cover all the domain relevant to NN-SRC ($F=R=1$) 
then $P_{pn/pX} = 1$.

The function  $R$ can be estimated in the NN-SRC model in which 
the decay function, $D^{pn}$ is a convolution of 
two density matrices representing the 
relative~($\rho^{pn}_{SRC}$) and c.m.~($\rho^{pn}_{c.m.}$)  momentum 
distributions of the $pn$-SRC: 
\begin{eqnarray}
D^{pn}  & =  & \rho^{pn}_{SRC}(\alpha_{rel},\vec p_{t,rel})\cdot
\rho^{pn}_{c.m.}(\alpha_{c.m.},\vec p_{t,c.m.}){\alpha_n\over
\alpha_{c.m.}}\times
\nonumber \\
& & \delta\left(P_{R+}- {m^2+p_{t,n}^2\over m\alpha_n} - {M^2_{A-2}+
p^2_{t,c.m.}\over m(A-\alpha_{c.m.})}\right),
\label{D_c.m.}
\end{eqnarray}
where $\alpha_{rel} = {\alpha-\alpha_n\over \alpha_{c.m.}}$, 
$p_{t,rel} = p_t - {\alpha\over \alpha_{c.m.}}p_{tn}$, 
$\alpha_{c.m.} = \alpha+\alpha_n$, and $p_{t,c.m.} = p_t + p_{tn}$. 
Within  the SRC model~\cite{FSreps}, $\rho^{pn}_{SRC}$ is related 
to the LC density matrix of the deuteron as:
\vspace{-0.2cm}
\begin{eqnarray}
 \rho_{SRC}(\alpha,p_t) = a_{pn}(A){\Psi^2_D(k)\over 2-\alpha}\sqrt{m^2+k^2}
\label{rc.m.}
\end{eqnarray}
where  $\Psi_D(k)$  is the deuteron wave function, and 
$k = \sqrt{{m^2+p^2_t\over \alpha(2-\alpha)} - m^2}$ ($0< \alpha < 2$). The 
parameter $a_{pn}(A)$ is the probability (relative to the deuteron) of
having 
a $pn$ SRC pair in nucleus $A$. 
Note that it cancels out in 
$R$, while $P_{pn/pX}$ 
contains information on $a_{pn}/a_{NN}$, with $a_{NN}$ related to 
the probability of NN-SRCs.

The c.m. motion of the SRC relative to the $(A-2)$ spectator system is 
described  by a Gaussian ansatz similar to Ref.\cite{CS96} with $\sigma$ 
being a parameter.
This distribution can be expressed through 
the LC momentum of the c.m. of the SRC as follows:
\vspace{-0.2cm}
\begin{equation}
\rho_{c.m.}(\alpha,p_{t}) = 2m\left({1\over 2\pi \sigma^2}\right)^{3\over 2}
e^{-{m^2(2-\alpha)^2+p_{t}^2\over 2\sigma^2}}.
\end{equation}
It is normalized as $\int \rho_{c.m.}(\alpha,p_{t}) {d\alpha\over
\alpha}d^2p_t = 1$.

The factorization of Eq.(\ref{D_c.m.}) is specific to the NN-SRC model, in 
which  it is assumed that the singular character of the NN potential at
short 
distances defines the main structure of the nucleon momentum distribution, 
and that it is weakly affected by the interaction of the NN-SRC  with the 
$A-2$ nuclear system. This approach gives a good description of the
spectral function of $^3He$\cite{CFSS91}, as well as medium
nuclei\cite{CS96}, 
with fitted values of total probabilities for NN-SRC which are up to 
$20\%$ less than the values extracted in Ref.\cite{Kim1,Kim2,Day}.

\begin{figure}[ht]
\vspace{-0.4cm}
\centerline{\epsfig{height=4cm,width=8cm,file=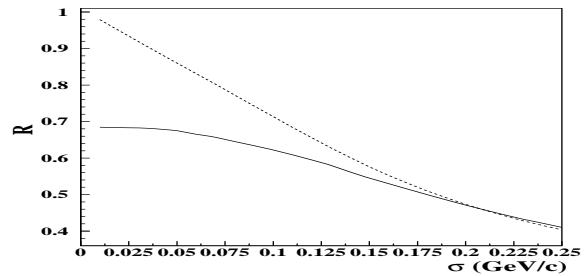}}
\vspace{-0.4cm}
\caption{ $R$ as a function of $\sigma$. The dashed line is for the same
cuts 
applied to both the $(p,pp$) and $(p,ppn)$ reactions; the solid line is
for the cuts of the EVA experiment.}
\vspace{-0.4cm}
\end{figure}
Figure~2 shows $R$ as a function of $\sigma$, calculated  within the above 
approach, as a dashed line. Here we used the same $0.6< \alpha < 1.1$  
and $k_F<p_{t}<0.55$~GeV/c cuts in both $(p,pp)$ and $(p,ppn)$ reactions.
It can be seen from Fig.2 that $R$ approaches unity at 
$\sigma\rightarrow 0$ which is consistent with the prediction of the SRC
model 
with the c.m. at rest. Figure~2 shows a large sensitivity of $R$ to the 
width of the c.m. momentum  distribution.

To extract $P_{pn/pX}$  from the data of Ref.\cite{Eip1,Eip2} using 
Eq.(\ref{PpnSRC})
one needs estimates of $R$, $T_n$, and $F$, for the kinematic cuts 
of the experiment, which are:
\vspace{-0.25cm}
\begin{eqnarray}
\mbox{struck proton: }&&  0.6 < \alpha < 1.1;  \  p_2 > p^{min}    
\nonumber \\ 
\mbox{recoil neutron: }&& 0.9 < \alpha_n < 1.4;  \ p^{min}< p_n < 0.55
\nonumber \\
&&  72^0 < \theta_n < 132^0.
\vspace{-0.4cm}
\label{cuts}
\end{eqnarray}
Here we follow Ref.\cite{Kim2} and set $p^{min}=0.275$~GeV/c  to
suppress mean field 
contribution to $\le 1\%$ 

{\boldmath $R$:}
The solid curve in Fig.2 represents $R$ calculated for the experimental 
cuts of Eq.(\ref{cuts}). It is different from unity even for 
$\sigma\rightarrow 0$ due to the different integration ranges for 
$p_2$ and $p_n$; Eq.(\ref{cuts}) places no restriction 
on $p_{2t}\equiv p_t$, while $p_{nt} \ge 204$~MeV/c.

{\boldmath $T_n$:} We used $T_n =0.85$ obtained from the 
simulation based on  Eq.(\ref{p2pn}) using the experimental limits of 
Eq.(\ref{cuts}). $T_n$ includes  Pauli Blocking effects according to 
Ref.\cite{PP} as well as a small correction due to neutrons from the 
rescattering of the fast incoming and outgoing protons calculated in the 
eikonal approximation\cite{Yael,FPSS}. 

{\boldmath $F$:}
The ratio $F$ is taken from Ref.\cite{Eip2},  where it 
is quoted  as $F =0.49\pm 0.13$ for $p^{min}=0.22$~GeV/c. For 
$p^{min}=0.275$~GeV/c  our analysis of the same data gives:
\vspace{-0.2cm}
\begin{equation}
\vspace{-0.2cm}
F= 0.43_{-0.07}^{+0.11}\cdot
\label{fexp}
\end{equation}
The uncertainty is dominated by the low statistics. Since the lower limit of
the extracted 
$P_{pn/pX}$ 
is sensitive to the lower limit on $F$ (Eq.(\ref{PpnSRC})) we choose 
for  $F$ the 
best one standard deviation low limit allowed by the data.

Using the above values of $R$, $T_n$, and $F$, we estimate 
$P_{pn/pX}$ from  Eq.(\ref{PpnSRC}). 
Figure~3 shows the $\sigma$ dependence of $P_{pn/pX}$ for 
$F=0.36$, $0.43$ and $0.55$, respectively.
Since $P_{pn/pX}\le 1$, there is an 
interesting correlation  between  $\sigma$ and $P_{pn/pX}$, which allows us 
to put a constraint on  $\sigma$. For example for 
$F=0.43$, $\sigma$ can not exceed $174$~MeV/c. To evaluate  $P_{pn/pX}$
we use 
the  magnitude of $\sigma^{exp} = 143\pm17$~MeV/c  
extracted from the same data set~\cite{Eip2}.  
\begin{figure}[ht]
\vspace{-0.4cm}
\centerline{\epsfig{height=5.0cm,width=8cm,file=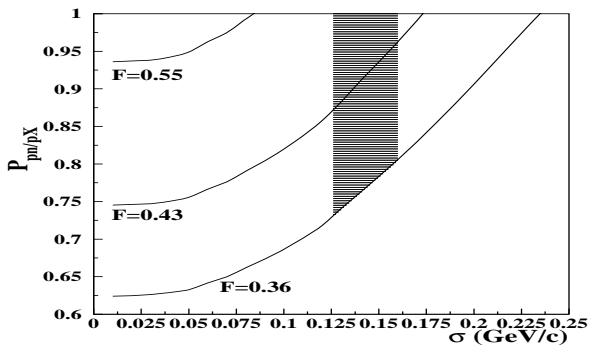}} 
\vspace{-0.4cm}
\caption{$P_{pn/pX}$ as a function of $\sigma$. The shaded area corresponds 
to the $P_{pn/pX}$ values at $\sigma_{exp}=143\pm 17$~MeV/c.} 
\vspace{-0.2cm}
\end{figure}
This value is in excellent agreement with 
the theoretical expectation of $139$~MeV/c of Ref.\cite{CS96}. 
Note that $\sigma^{exp}$  dictates  the range of possible 
values for $P_{pn/pX}$.  From the  central~($0.43$) 
and minimal~($0.36$) values of $F$ we obtain:
\vspace{-0.3cm}
\begin{equation} 
P_{pn/pX} = 0.92^{+0.08}_{-0.18}.  
\label{P_pn_exp}
\vspace{-0.1cm}
\end{equation}
This result indicates that at least  $74\%$ of the time the removal of a 
fast proton  is accompanied  by  the emission of a fast recoil neutron. 
It allows us also to estimate an upper 
limit of the ratio of absolute probabilities of $pp-$ to $pn-$ 
SRCs ($P_{pp}\over P_{pn}$), by assuming that 
$pp$-SRCs carry all the remaining strength in ``pX''
configuration\cite{comment}:
\vspace{-0.4cm}
\begin{equation}
{P_{pp}\over P_{pn}} 
\le {1\over 2} (1-P_{pn/pX})= 0.04_{-0.04}^{+0.09}.
\vspace{-0.4cm}
\label{P_pp_pn}
\end{equation}
This result can be used to estimate separately the  absolute
probabilities of $pn$, 
$pp$ and $nn$ SRCs in the nuclear wave function. For this we use 
the total probability of NN-SRCs ($P_{NN}(^{12}C) = 0.20 \pm 0.042$)  
obtained  by combining the results of large $Q^2$ and $x_B$  inclusive 
$A(e,e')X$ data from  Refs.\cite{Day,Kim1,Kim2}.  
The value of $P_{NN}$ is practically  independent of  
whether or not we account for  the c.m. motion of the SRC.
Using $P_{NN}$ together with Eq.(\ref{P_pp_pn}) we obtain
$P_{pn}(^{12}C) = 0.184 \pm 0.045$ and $P_{pp}(^{12}C) = P_{nn}(^{12}C)
\leq 0.03$.

Summarizing, from  Eqs.(\ref{P_pn_exp},\ref{P_pp_pn}) we conclude 
that  our analysis of $(p,pp)$ and $(p,ppn)$ reactions 
indicates  that if a nucleon with momentum between $275-550$~MeV/c is 
removed from the nucleus using a high momentum and energy transfer
probe, at least 74$\%$ of the time it will originate from $pn$-SRCs. 
The data also show  significantly  smaller (a factor of 6 at least) 
probabilities for $pp$ and $nn$ NN-SRCs as compared to 
$pn$ NN-SRCs in Carbon.
This result may have an important implications for modeling the  
equation of state of asymmetric nuclear matter. It indicates that the 
equation of state of neutron stars could  be affected 
by the large suppression of the number of 
neutrons above the Fermi sea. 
Future experiments involving high-$Q^2$ $(e,e'pp)$ 
and $(e,e'pn)$ reactions off medium-to-heavy nuclei (e.g.,
Ref.\cite{ebww}), 
as well as theoretical studies based upon realistic
treatments of SRCs and final-state-interaction
effects (see, e.g., Ref.\cite{refadd1,refadd2,refadd3}), will allow one to 
verify and significantly improve the accuracy of the present results.

We would like to acknowledge the contribution of the EVA collaboration at 
BNL as well as useful discussions with J.~Alster, K.~Egiyan and A.~Gal. 
This research is supported by the Israel Science Foundation, the US-Israeli 
Binational Scientific Foundation, the US Department of Energy and the
National Science Foundation.

\end{document}